\documentclass[prd,showpacs,amsmath,showkeys,twocolumn,floatfix,amssymb, preprintnumbers, nofootinbib, superscriptaddress]{revtex4} 
\usepackage{epsfig,dcolumn}
\usepackage{graphicx}
\usepackage{comment} 
\usepackage[usenames]{color}
\usepackage{graphicx}
\usepackage{bm}
 \usepackage{ifpdf}

\def\lsim{\mathrel{\rlap{\lower4pt\hbox{\hskip1pt$\sim$}}
    \raise1pt\hbox{$<$}}}
\def\gsim{\mathrel{\rlap{\lower4pt\hbox{\hskip1pt$\sim$}}
    \raise1pt\hbox{$>$}}}
\begin{document}

\title{  The three-particle system on a torus}

\author{Peng~Guo}
\email{pguo@jlab.org}
\affiliation{Thomas Jefferson National Accelerator Facility, 
Newport News, VA 23606, USA}

\preprint{JLAB-THY-13-1700 }

\date{\today}

\begin{abstract} 
 Based on Lippmann-Schwinger equation approach, we discuss a three-particle system in finite volume.  A set of equations which relate the discrete finite-volume energies to the scattering amplitudes are derived under the approximation of the isobar model relevant for the case of narrow two- and three-body resonances.     
  \end{abstract} 


\maketitle


\section{Introduction}
\label{intro} 

Recent developments in the application of variational methods \cite{Michael:1985ne,Luscher:1990ck,Blossier:2009kd} to large bases of hadron interpolating fields have made the extraction of the excited spectrum of hadronic states a realistic possibility (see e.g. \cite{Jo:2010,Edwards:2011, Jo_scatt:2012}).  Moreover, these new developments in lattice QCD  provide a solid  foundation for studying multi-hadron scattering in lattice QCD \cite{Feng:2011,Jo_scatt:2011,Jo_scatt:2012,Beane_scatt:2012,Lang_scatt:2011, Aoki:2011yj,Dudek:2012xn}.   Because lattice QCD is formulated in Euclidean space, we do not have direct access to scattering amplitudes \cite{Maiani:1990ca}. Fortunately, in a finite volume, L\"uscher formula  \cite{Lusher:1991} presents a way to  relate the discrete  energy spectrum in lattice QCD  to  scattering amplitudes. Since then,   the framework derived by L\"uscher in \cite{Lusher:1991} has been extended to moving frames \cite{Gottlieb:1995,Lin:2001,Christ:2005,Bernard:2007,Bernard:2008},  and  the  inelastic region by including coupled-channel effects \cite{Liu:2005,Doring:2011,Aoki:2011,Briceno:2012yi, Hansen:2012tf,Guo:2013cp}.

Based on the approach   developed in \cite{Guo:2013cp} for the scattering of two-particle system in finite volume, in this work,  we discuss   the three-particle system, considering the finite-volume representation of the isobar model \cite{Goradia:1977,Ascoli:1975} in which interactions in the three-particle system are approximated by two-body scattering. We derive   expressions which relate finite-volume energy shifts to isobar-model production amplitudes.

The paper is organized as follows.   In Section \ref{threebody} we discuss the three-particle system in finite volume. The summary and outlook are given in Section \ref{summary}.


\section{Three-particle scattering on a torus within an isobar approximation}\label{threebody}
In QCD, because of the low mass of pions, three-particle states become important even at relatively low energies. Methods for extracting three-particle amplitudes from lattice simulations are thus  needed.   In general, the scattering of a multiple-particle system is quite complicated, however, in some cases the quasi-two body approximation has proven to be quite successful \cite{Faddeev:1960,Faddeev:1965}. A recent consideration of the three-particle system on a torus in the framework of Faddeev equations is presented in \cite{Rusetsky:2012}.

\begin{figure*}
\includegraphics[width=0.99\textwidth]{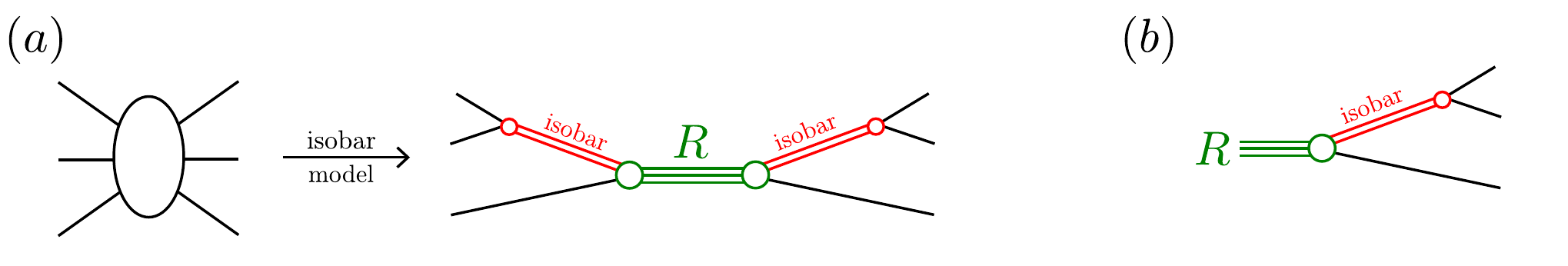}
\caption{ A schematic view of the isobar approximation to three-particle scattering. (a) The general $3\to3$ amplitude is assumed to be dominated by two-particle `isobar' resonances and to pass through a sharp three-particle resonance, $R$. (b) We assume the relevant singularities of the scattering amplitude are captured by considering the decay of the sharp intermediate resonance.
\label{3p}}
\end{figure*}

In this work, we are interested in establishing the presence of three-particle and two-particle resonances in a three-particle final-state system. In lattice QCD, the multi-hadron interpolating fields operators $\mathcal{O}$ are related to the total energies of multi-hadron system  through the correlation function $C(t) =  \langle 0 | \mathcal{O}(t)  \mathcal{O}^{\dag}(0) | 0\rangle $  by relation
\begin{eqnarray}
C(t) = \sum_{n} \langle 0 | \mathcal{O}  | n \rangle \langle n | \mathcal{O}^{\dag}  | 0\rangle e^{- E_{n} t}.
\end{eqnarray}
Therefore, instead of building the relations between three-to-three particles scattering amplitudes and $E_{n}$ in finite volume in \cite{Briceno:2012},   we intend to reconstruct a three-body resonance by its decay products, see figure \ref{3p}.
When two-body resonances occur, the decay or production of a three-particle final state system can be \emph{approximately} described in terms of isobars recoiling from a bachelor particle\cite{Goradia:1977,Ascoli:1975,Hwa:1963,Aaron:1973}. As an example, the decay process of the strange axial meson resonance $K_{1}(1400)$ to the $K \pi \pi$ final state can be quite well approximated by $K^{*}(892)\, \pi$ in $S$-wave, with the $K^*$ isobar decaying to $K \pi$ in a $P$-wave.

As presented in Appendix \ref{production}, we consider the intermediate resonance to be effectively an asymptotic state and parameterize its decay via a two-body isobar resonance recoiling from a spectator particle.

In this section, we will generalize the  methods developed in \cite{Guo:2013cp}   to apply to the non-relativistic three-particle system in the isobar approximation. For simplicity  we consider three scalar particles with equal masses, $m$. As discussed in previous paragraph, we consider the production of a three-particle final state from the decay of a sharp resonance, $| R  \rangle$, which we will treat as an asymptotic state whose origin is in the short-distance $q \bar{q}$ dynamics, rather than the final-state interactions. 
We proceed assuming that the dynamics of transition from a resonance state in the far past to a three-particle state in the far future is described by a three-particle final-state interaction Hamiltonian $ H^{\textsc{FSI}} =H^{\textsc{0}} + V^{\textsc{FSI}} $. Under the assumption of strong pair-wise final state interactions between two particles, strong enough to generate a two-body isobar resonance, the final state interaction Hamiltonian of the three-particle system takes the form 
\begin{eqnarray}\label{hamiltonianFSI}
 H^{\textsc{FSI}} =\sum_{k=1,2,3}    \frac{ \mathbf{ p}_{k}^{2}}{2m} + \sum_{k=1,2,3}^{i \neq j \neq k} \tilde{V}(\mathbf{ x}_{i}- \mathbf{ x}_{j}), \nonumber
\end{eqnarray}
where $(\mathbf{ p}_{k},\mathbf{ x}_{k})$ are the momentum and coordinate of the  $k^\mathrm{th}$ particle.

The amplitude to produce the three-particle final state $|\mathbf{ p}_{1}  \mathbf{ p}_{2}   \mathbf{ p}_{3} \rangle$ is defined by
\begin{eqnarray} 
   f (  \mathbf{ p}_{1}  \mathbf{ p}_{2}   \mathbf{ p}_{3}   \leftarrow R)  =  \langle  \mathbf{ p}_{1}  \mathbf{ p}_{2}   \mathbf{ p}_{3}  |  V^{\textsc{FSI}}   |\Psi^{\mathrm{in}} (0) \rangle,  
\end{eqnarray} 
where the incoming wave state at $t=0$ is given by $ |  \Psi^{\mathrm{in}} (0) \rangle  =  U(0, -\infty) | R\rangle$ and     $U(t, t_{0}) =\mathcal{T}  \left \{ \exp \left [- i  \int_{t_{0}}^{t} dt'  e^{i H^{\textsc{0}}t'  }  V^{\textsc{FSI}}  e^{-i  H^{\textsc{0}} t'  }    \right ]  \right \}$. The   wavefunction of three final state  particles     in coordinate space is given by 
 \begin{eqnarray}
  \psi_{JM}(\mathbf{ x}_{1}, \mathbf{ x}_{2}, \mathbf{ x}_{3}) = \langle \mathbf{ x}_{1}  \mathbf{ x}_{2}   \mathbf{ x}_{3}     |\Psi^{\mathrm{in}} (0) \rangle.
  \end{eqnarray} 
Under the isobar model approximation, where $ V^{\textsc{FSI}}$ includes only pair-wise interactions and where rearrangement among different isobar pairs is ignored, the three-particle production amplitude  is approximated by
 \begin{eqnarray} 
   f (  \mathbf{ p}_{1}  \mathbf{ p}_{2}   \mathbf{ p}_{3}   \leftarrow R)   \approx  \sum_{k=1,2,3}^{ i \neq j \neq k} \langle  \mathbf{ q}_{ij}   \,   \mathbf{ k}_{k} | t^{(ij)}    |R \rangle, 
\end{eqnarray} 
where $t^{(ij)}$ is the two-body $T$-matrix operator for pair $(ij)$, the relative momenta $( \mathbf{ q}_{ij}   \, ,  \mathbf{ k}_{k} )$ are given by $  \mathbf{ q}_{ij} =\frac{1}{2} (  \mathbf{ p}_{i}  -  \mathbf{ p}_{j} ) $ and $   \mathbf{ k}_{k} =\frac{1}{3} ( \mathbf{ p}_{i}  +  \mathbf{ p}_{j}- 2  \mathbf{ p}_{k} ) $ respectively. The $S$-matrix describing the transition from $|R\rangle$ to  $|\mathbf{ p}_{1}  \mathbf{ p}_{2}   \mathbf{ p}_{3} \rangle$ is discussed further in Appendix \ref{production}.

 In the center-of-mass frame, the  wavefunction of the three-particle system has the following form \cite{Faddeev:1960}
\begin{eqnarray}\label{three_tot_wave}
\psi_{JM}(\mathbf{ x}_{1}, \mathbf{ x}_{2}, \mathbf{ x}_{3}) = \sum^{i\neq j \neq k}_{k=1,2,3}\phi^{(ij)}_{JM}(\mathbf{ r}_{ij}, \mathbf{ r}_{k} ) ,  \nonumber
\end{eqnarray}
where $\mathbf{ r}_{ij}=\mathbf{ x}_{i}-\mathbf{ x}_{j}$ is the relative position between the two particles forming the $(ij)$ isobar and  $\mathbf{ r}_{k}=\frac{1}{2}(\mathbf{ x}_{i}+\mathbf{ x}_{j}) -\mathbf{ x}_{k}$ is the relative position of the spectator particle and the isobar. The solution of the  Schr\"odinger equation $H^{\textsc{FSI}} \, \psi_{JM} = E\, \psi_{JM}$
is given by 
\begin{align} 
 \phi_{JM}^{(ij)}(\mathbf{ r}_{ij}, \mathbf{ r}_{k})  &= \int\! d^{3} \mathbf{ r}'_{ij}\, d^{3} \mathbf{ r}'_{k}  \;
  G_{0}(  \mathbf{ r}_{ij} -\mathbf{ r}_{ij}' , \mathbf{ r}_{k}-\mathbf{ r}_{k}';E) \nonumber \\
 &\quad\quad\times    \tilde{V}_{ij}(\mathbf{ r}_{ij}')  \,  \psi_{JM}( \mathbf{ x}_{1}',  \mathbf{ x}_{2}',\mathbf{ x}_{3}')    , \nonumber
\end{align}
 with  $G_0$ being the  non-relativistic three-body Green's function, 
\begin{eqnarray}
 G_{0}(  \mathbf{ r}_{ij}, \mathbf{ r}_{k};E)  =    \int\!  \frac{d^{3} \mathbf{ q}_{ij}}{(2\pi)^{3}}  \, \frac{d^{3} \mathbf{ k}_{k}}{(2\pi)^{3}} \;  \frac{e^{i \mathbf{ q}_{ij} \cdot  \mathbf{ r}_{ij}   +i \mathbf{ k}_{k} \cdot  \mathbf{ r}_{k}   } }
 {E - 3m -  \frac{\mathbf{ q}_{ij}^2  }{m}  - \frac{3 \mathbf{ k}_{k}^{2}}{4m}  }  .   \nonumber \\
 \end{eqnarray}
 Using the identity
  \begin{align} 
 &\frac{1}{E-3 m - \frac{\mathbf{ q}_{ij}^{2}}{m} - \frac{3 \mathbf{ k}_{k}^{2}}{4m}} = \frac{i}{2\pi} \int_{-\infty}^{\infty}  d E_{ij}    \nonumber \\
 &\times \frac{1}{E-m -E_{ij} -\frac{3 \mathbf{ k}_{k}^{2}}{4m} + i0} \frac{1}{ E_{ij} -2 m- \frac{\mathbf{ q}_{ij}^{2}}{m} + i0 }  . \nonumber \\
\end{align} 
We may represent the  three-body Green's function by a product of two independent two-body  Green's functions 
  \begin{align}
G_{0}(  \mathbf{ r}_{ij},& \mathbf{ r}_{k};E)   
=  \frac{i}{2 \pi }   \int_{ - \infty}^{\infty}  \!\!\!\!d E_{ij}   \, G^{ (ij)+k}_{0}(  \mathbf{ r}_{k};E)    \, G^{  i+j}_{0}(  \mathbf{ r}_{ij};E_{ij})    ,  \nonumber
 \end{align}
 where 
   \begin{eqnarray}
 G^{ (ij)+k}_{0}(  \mathbf{ r}_{k};E)   =   \int \!     \frac{d^{3} \mathbf{ k}_{k}}{(2\pi)^{3}}    \;\frac{ e^{ i \mathbf{ k}_{k} \cdot  \mathbf{ r}_{k}   } }{E -m - E_{ij}  - \frac{ 3\mathbf{ k}_{k}^{2}}{4 m}     }  \nonumber
 \end{eqnarray}
 is the free Green's function for propagation of the system, of total energy $E$, made of the isobar pair $(ij)$ (with invariant mass $E_{ij}$) plus the spectator particle,  and where
   \begin{eqnarray}
 G^{  i+j}_{0}(  \mathbf{ r}_{ij};E_{ij})   =    \int \! \frac{d^{3} \mathbf{ q}_{ij}}{(2\pi)^{3}}   \;  \frac{e^{i \mathbf{ q}_{ij} \cdot  \mathbf{ r}_{ij}      }}{ E_{ij} - 2m  - \frac{ \mathbf{ q}^{2}_{ij}}{m}} , \nonumber
 \end{eqnarray}
is the free Green's function describing the propagation of particles, $(i,j)$, inside  the isobar.

The Green's functions  are explicitly given by 
 \begin{align}\label{three_green}
 G_{0}(  \mathbf{ r}_{ij},& \mathbf{ r}_{k};E)  \rightarrow  \frac{i}{2\pi  } \int_{2m}^{E- m}  \!\!\!\!\!\!\!\!d E_{ij} \,   \frac{4m^{2}}{3(4\pi)^2}   \frac{e^{i q_{E_{ij}} r_{ij}}}{r_{ij}}  \frac{e^{i k_{E_{ij}} r_{k}}}{r_{k}},  \nonumber \\
 \end{align}
with 
  \begin{eqnarray}
 q_{E_{ij}} &=&  \sqrt{m (E_{ij} - 2 m)},   \nonumber \\
 k_{E_{ij} }&=& \sqrt{ \frac{4}{3}m (E-m - E_{ij})  }. \nonumber
 \end{eqnarray} 
 In  Eq.(\ref{three_green}),   $ \int_{-\infty}^{\infty} d E_{ij}$ has been replaced by $ \int_{2m}^{E- m} d E_{ij}$, because the  three-body Green's function has   oscillatory  behavior  for $E_{ij}$ only inside  the physical region.   The 
   three-particle wavefunction is given by 
\begin{align} 
\psi_{JM}&(\mathbf{ x}_{1}, \mathbf{ x}_{2}, \mathbf{ x}_{3})   \nonumber \\
 &\rightarrow  \frac{i}{2\pi  } \sum_{k=1,2,3}     \int_{2m}^{E-m} \!\!\!\!\!\!\!\!d E_{ij}  \!\!\!\sum_{\substack{ S_{ij} M_{S_{ij}} \\L_{k} M_{L_{k}} } }   \nonumber \\
  &\quad\times  (  i q_{E_{ij}} )\, h^{+}_{S_{ij}}(q_{E_{ij}} r_{ij})\,  i^{S_{ij}}\, Y_{S_{ij} M_{S_{ij}}} ( \mathbf{ \hat{r}}_{ij})   \nonumber \\
 &\quad\times (i k_{E })\,  h^{+}_{L_{k}} (k_{E_{ij} } r_{k})\, i^{L_{k}}\, Y_{L_{k} M_{L_{k}}} (\mathbf{ \hat{r}}_{k})  \nonumber \\
 &\quad\times  f^{(ij)}_{JM;S_{ij} M_{S_{ij}};L_{k} M_{L_{k}} } (q_{E_{ij}} , k_{E_{ij} } ) ,  \nonumber
\end{align}
where the   three-body production amplitudes are defined by
\begin{align}
f&^{(ij)}_{JM;S_{ij} M_{S_{ij}};L_{k} M_{L_{k}} }  (q_{E_{ij}} , k_{E_{ij} } )   \nonumber \\
  &\quad=     \frac{  4 m^{2}}{3}   \int\! d^{3} \mathbf{ r}'_{ij}\, d^{3} \mathbf{ r}'_{k}    \nonumber \\ 
 &\quad\quad\times  i^{- S_{ij}} \, j_{S_{ij}} (q_{E_{ij}}    r'_{ij})\,  Y^{*}_{S_{ij} M_{S_{ij}}} ( \mathbf{ \hat{r}}_{ij}')  \nonumber \\
 &\quad\quad\times  i^{- L_{k}} \, j_{L_{k}} (k_{E_{ij} }    r'_{k}) \,  Y^{*}_{L_{k} M_{L_{k}}} ( \mathbf{ \hat{r}}_{k}')  \nonumber \\
  &\quad\quad\times  \tilde{V}_{ij}(\mathbf{ r}_{ij}')  \,  \psi_{JM}(  \mathbf{ x}_{1}',\mathbf{ x}_{2}' ,\mathbf{ x}_{3}')  . \label{three_amp}
\end{align}
The total angular momentum of the three-particle system is $J$, made up of $(S_{ij}, M_{S_{ij}})$, the spin and its $z$-axis projection of the isobar pair $(ij)$, and $(L_{k} , M_{L_{k}})$, the relative orbital angular momentum and projection between $k^\mathrm{th}$ particle and isobar pair $(ij)$. 

The partial wave production amplitude  
can be parametrized by
\begin{align}\label{three_pw}
&f^{(ij)}_{JM;S_{ij} M_{S_{ij}};L_{k} M_{L_{k}} }  (q_{E_{ij}} , k_{E_{ij} } )   \nonumber \\
 & \quad =  \langle S_{ij} M_{S_{ij}} ; L_{k} M_{L_{k}} | JM  \rangle\,    f_{S_{ij}} (q_{E_{ij}}) \, f_{ S_{ij}L_{k} J} (k_{E_{ij}}), 
\end{align}
where 
\begin{equation} 
f_{S_{ij}}(q_{E_{ij}}) =\frac{4\pi}{q_{E_{ij}}  } e^{i \delta_{S_{ij}}  } \sin \delta_{S_{ij}} ,\nonumber
\end{equation} 
 is the partial wave scattering amplitude of two spinless particles inside the isobar pair $(ij)$
  and 
$ f_{S_{ij}L_{k}J} (k_{E_{ij}})$   is the partial wave production amplitude of the isobar pair $(ij)$  and the spectator particle.

\begin{widetext}

The final expression for the three-particle production wavefunction, including the homogeneous term, reads
\begin{align} 
\psi_{JM}(\mathbf{ x}_{1}, \mathbf{ x}_{2}, \mathbf{ x}_{3})    \rightarrow \frac{i}{2\pi  } \sum_{k=1,2,3}       &\sum_{\substack{S_{ij} M_{S_{ij}}\\L_{k} M_{L_{k}}}}  \langle S_{ij} M_{S_{ij}} ; L_{k} M_{L_{k}} | JM  \rangle  \, Y_{S_{ij} M_{S_{ij}}} ( \mathbf{ \hat{r}}_{ij})  \, Y_{L_{k} M_{L_{k}}} (\mathbf{ \hat{r}}_{k}) \nonumber \\
 &\times \int_{2m}^{E-m} \!\!\!\!\!\!\!\! d E_{ij}  \; 
  i^{S_{ij}}  \left[(4\pi) \,  j_{S_{ij}} (q_{E_{ij}} r_{ij})+i \,q_{E_{ij}} \, h^{+}_{S_{ij}}(q_{E_{ij}} r_{ij})  \,  f_{S_{ij}} (q_{E_{ij}} ) \right] \nonumber \\
 &\hspace{1.5cm} \times i^{L_{k}} \left[ (4\pi)\,   j_{L_{k}} (k_{E_{ij}} r_{k} )+ i \,k_{E_{ij}} \, h^{+}_{L_{k}} (k_{E_{ij}} r_{k}) \, f_{S_{ij}L_{k}J} (k_{E_{ij}}) \right] .  \nonumber 
\end{align}
  Boosting the three-particle system from the center-of-mass frame to a lab frame with  the   center-of-mass of isobar pair $(ij)$ fixed at the origin, $\mathbf{ x}_{i}+\mathbf{ x}_{j}=\mathbf{ 0}$, the wavefunction of  the three-particle system  at the lab frame   can be  written as the product of  a plane-wave,   $e^{i \mathbf{ P} \cdot (\mathbf{ x}_{1}+\mathbf{ x}_{2}+\mathbf{ x}_{3})/3 } $,  and  a piece  depending only on relative coordinates,  $\mathbf{ r}_{ij}=\mathbf{ x}_{i}-\mathbf{ x}_{j}$ and $ \mathbf{ r}_{k}= -\mathbf{ x}_{k}$. 
  As in the case of two-particle  scattering   \cite{Gottlieb:1995},   requiring periodicity of the lab frame  wavefunction   with respect to    $\mathbf{ r}_{ij} $ and $ \mathbf{ r}_{k} $, we find that the boundary condition of CM frame wavefunction reads  
\begin{eqnarray}\label{three_boundary}
  \psi^{(L)}_{JM}(  \mathbf{ r}_{ij}+\mathbf{ n}_{ij} L,\mathbf{ r}_{k}+\mathbf{ n}_{k} L )  = e^{   i \mathbf{ Q}  \cdot  \mathbf{ n}_{k}    L  } \,  \psi^{(L)}_{JM}(  \mathbf{ r}_{ij},\mathbf{ r}_{k}) ,  \nonumber
\end{eqnarray}
  where $k=1,2,3$ and  $\mathbf{ Q} $ is the Bloch wave vector for the three-particle system.
   The connection of  the Bloch wave vector $\mathbf{ Q}  $ to the total momentum of the  three-particle system is given by      
   $\mathbf{ P}  = 3\gamma \mathbf{ Q}   $.     Using the periodicity of the potential $ \tilde{V}_{ij}(\mathbf{ r}_{ij}'+ \mathbf{ n}_{ij} L ) =  \tilde{V}_{ij}(\mathbf{ r}_{ij}' )$,  the three-particle Lippmann-Schwinger equation on a torus can be written 
\begin{align}
\psi^{(L,\mathbf{ Q})}_{JM}&(\mathbf{ x}_{1}, \mathbf{ x}_{2}, \mathbf{ x}_{3})     = \sum_{k=1,2,3}     \int_{L^{3}} \!\!\!d^{3} \mathbf{ r}'_{ij}  \int_{L^{3}}  \!\!\!d^{3} \mathbf{ r}'_{k} \, G_{\mathbf{ Q}}(  \mathbf{ r}_{ij} -\mathbf{ r}_{ij}'  , \mathbf{ r}_{k}-\mathbf{ r}_{k}'  ;E)   \,    \tilde{V}_{ij}(\mathbf{ r}_{ij}' ) \,  \psi^{(L,\mathbf{ Q})}_{JM}(  \mathbf{ x}'_{1},  \mathbf{ x}'_{2} ,\mathbf{ x}'_{3} )    ,\nonumber
\end{align}
 where the periodic three-body Green's function is given by
\begin{align}
 G_{\mathbf{ Q}}&(  \mathbf{ r}_{ij}, \mathbf{ r}_{k};E)   = \sum_{\mathbf{ n}_{ij}, \mathbf{ n}_{k} \in \mathbb{Z}^{3}} \!\! G_{0}(  \mathbf{ r}_{ij}- \mathbf{ n}_{ij} L, \mathbf{ r}_{k} -  \mathbf{ n}_{k} L;E)   \,    e^{ i   \mathbf{ Q}  \cdot   \mathbf{ n}_{k}  L } ,   \nonumber 
 \end{align}
or asymptotically  
   \begin{align}
G_{\mathbf{ Q}}&(  \mathbf{ r}_{ij}, \mathbf{ r}_{k};E)     \to    \frac{i}{2\pi  } \int_{2m }^{E-m} \!\!\!\!\!\!\!\!d E_{ij}   \frac{4m^{2}}{3}   \,  \frac{1}{L^{3}} \!\!\!\sum_{  \mathbf{ q}_{ij} \in \mathbf{ P}_{\mathbf{ 0}}  }    \frac{e^{i  \mathbf{ q}_{ij} \cdot   \mathbf{ r}_{ij}} }{ q_{E_{ij}}^{2}- \mathbf{ q}_{ij}^{2}       }  \frac{1}{ L^{3}} \!\!\! \sum_{   \mathbf{ k}_{k} \in \mathbf{ P}_{  \mathbf{ Q}}  }    \frac{e^{i  \mathbf{ k}_{k} \cdot   \mathbf{ r}_{k}} }{ k_{E_{ij}}^{2}- \mathbf{ k}_{k}^{2}       }, \nonumber
 \end{align}
where $\mathbf{ P}_{\mathbf{ Q}} = \{ \mathbf{ q} \in \mathbb{R}^{3}| \mathbf{ q}= \frac{2\pi}{L}\mathbf{ n} + \mathbf{ Q}, \mbox{ for }   \mathbf{ n}\in \mathbb{Z}^{3} \} $. Next, we use the expansion of the Green's function in  Eq.(B2) in \cite{Guo:2013cp}   and the definition of the scattering amplitude in Eq.(\ref{three_amp}) and (\ref{three_pw}), to obtain 
\begin{align} 
 \psi^{(L,\mathbf{ Q})}_{JM}&(\mathbf{ x}_{1}, \mathbf{ x}_{2}, \mathbf{ x}_{3})       \rightarrow  \frac{i}{2\pi  }  \sum_{k=1,2,3}     \sum_{\substack{S_{ij} M_{S_{ij}}\\L_{k} M_{L_{k}}}}  \sum_{\substack{S'_{ij} M'_{S_{ij}} \\  L'_{k} M'_{L_{k}}}}  \langle S_{ij} M_{S_{ij}} ; L_{k} M_{L_{k}} | JM  \rangle  \,  Y_{S'_{ij} M'_{S_{ij}}} ( \mathbf{ \hat{r}}_{ij})\,   Y_{L'_{k} M'_{L_{k}}} (\mathbf{ \hat{r}}_{k}) \nonumber \\
 &\quad\quad\times \int_{2m}^{E-m} \!\!\!\!\!\!\!\!d E_{ij}   \, 
      i^{S_{ij} } \, q_{E_{ij}}  \, f_{S_{ij}} (q_{E_{ij}}) \left[ \delta_{S_{ij} M_{S_{ij}}, S'_{ij} M'_{S_{ij}}} \, n_{S'_{ij}} (q_{E_{ij}} r_{ij})- \mathcal{M}^{(\mathbf{ 0})}_{S_{ij} M_{S_{ij}}, S'_{ij} M'_{S_{ij}}}  (q_{E_{ij}} )    \, j_{S'_{ij}} (q_{E_{ij}} r_{ij})    \right ]  \nonumber \\
 &\quad\quad\quad\quad\quad \quad\times i^{ L_{k}} \, k_{E_{ij}} \,  f_{ S_{ij}L_{k} J} (k_{E_{ij}} )  \left[ \delta_{L_{k} M_{L_{k}}, L'_{k} M'_{L_{k}}}  \,  n_{L'_{k}} (k_{E_{ij}} r_{k} )-\mathcal{M}^{(\mathbf{ Q})}_{L_{k} M_{L_{k}}, L'_{k} M'_{L_{k}}} (k_{E_{ij}}   )   \, j_{L'_{k}} (k_{E_{ij}} r_{k} )\right ] .  \nonumber
\end{align}

Matching a general wavefunction of   form  $\sum_{JM} c_{JM} \, \psi_{JM}(\mathbf{ x}_{1},\mathbf{ x}_{2},\mathbf{ x}_{3} )$  
with   $\sum_{JM} c_{JM} \, \psi^{(L,\mathbf{ Q})}_{JM}(\mathbf{ x}_{1},\mathbf{ x}_{2},\mathbf{ x}_{3} )$, 
 and projecting out the partial waves, neglecting rearrangement effects from crossed channels as appropriate in an isobar approximation, we get a set of equations  which have general form $ \int_{2m}^{E-m} d E_{ij}  \  F(E_{ij}, r_{ij}, r_{k})=0$. 
 Since  the variables $(r_{ij},r_{k})$ can be chosen arbitrarily, $    F(E_{ij}, r_{ij}, r_{k})=0$ must be satisfied for each $E_{ij}$. From this we obtain three determinant conditions, the first is 
 \begin{eqnarray}\label{three_body_cond_1}  
     \det \left [   \delta_{S_{ij} M_{S_{ij}} , S'_{ij} M'_{S_{ij}} } \,  \cot \delta_{S_{ij}} (q_{E_{ij}} )  -   \mathcal{M}^{(\mathbf{ 0})}_{S_{ij} M_{S_{ij}}, S'_{ij} M'_{S_{ij}}}  (q_{E_{ij}} ) \right ]  =0 ,
\end{eqnarray}
 which is Lu\"scher's formula for scattering between $i^\mathrm{th}$ and $j^\mathrm{th}$  particles inside the isobar pair $(ij)$. It provides the constraint  on the phase shifts $ \delta_{S_{ij}}$ as a function of  invariant mass of the isobar pair $(ij)$, $E_{ij}$. The second condition, 
 \begin{align}\label{three_body_cond_2} 
 \det    \bigg[   &\delta_{JM, J'M'} \, \delta_{L_{k}, L'_{k}} \,  \left [  \frac{1}{\frac{k_{E_{ij}} }{4\pi}  f_{S_{ij}L_{k} J}(k_{E_{ij}}) } + i \right  ]     \nonumber \\
 &\quad\quad -     \!\!\!\! \sum_{M_{S_{ij}} M_{L_{k}}  M'_{L_{k} }} \!\!\!\!\!\!\! \langle S_{ij} M_{S_{ij}} ; L'_{k} M'_{L_{k}} | J'M'  \rangle \,   \langle S_{ij} M_{S_{ij}} ; L_{k} M_{L_{k}} | JM  \rangle \,         \mathcal{M}^{ ( \mathbf{ Q})}_{L_{k} M_{L_{k}}, L'_{k} M'_{L_{k}}} (k_{E_{ij}}     )      \bigg] =0,  
\end{align}
  is a generalized Lu\"scher's formula in moving frames for production of   the spectator, $k^\mathrm{th}$ particle,  and  the isobar $(ij)$ with the specific spin $S_{ij}$ and mass $E_{ij}$ from initial state with  spin $J$. Thus,  Eq. (\ref{three_body_cond_2}) gives the constraint on  the production amplitudes  $ f_{S_{ij} L_{k} J}$  as   function of  the total energy, $E$,   for each individual partial wave of  isobar pair $(ij)$,  $S_{ij}$.  
  
  The final condition, 
 \begin{align}\label{three_body_cond_3} 
 \det    \bigg[ &\delta_{JM, J'M'}  \, \delta_{S_{ij}   , S'_{ij}   } \, \delta_{L_{k}, L'_{k}} \, \cot \delta_{S_{ij}} (q_{E_{ij}} ) \,   \left [  \frac{1}{\frac{k_{E_{ij}} }{4\pi}  f_{S_{ij}L_{k} J}(k_{E_{ij}}) } + i \right  ]    \nonumber \\
 &\quad-     \!\!\!\! \sum_{\substack{M_{S_{ij}} M'_{S_{ij}} \\ M_{L_{k}}  M'_{L_{k} }}} \!\!\! \langle S'_{ij} M'_{S_{ij}} ; L'_{k} M'_{L_{k}} | J'M'  \rangle  \,    \langle S_{ij} M_{S_{ij}} ; L_{k} M_{L_{k}} | JM  \rangle  \,  \mathcal{M}^{(\mathbf{ 0})}_{S_{ij} M_{S_{ij}}, S'_{ij} M'_{S_{ij}}}  (q_{E_{ij}} ) \,         \mathcal{M}^{ ( \mathbf{ Q})}_{L_{k} M_{L_{k}}, L'_{k} M'_{L_{k}}} (k_{E_{ij}}     )  \bigg] =0, \nonumber \\
\end{align}
 leads to an additional  constraint on both $ \delta_{S_{ij}}$ and  $ f_{S_{ij} L_{k} J}$ for the scattering between the $k^\mathrm{th}$ particle and all the allowed partial waves of  the isobar $(ij)$. 
 \end{widetext}  
 
 In the case that only a single partial wave $S_{ij}$ of the isobar pair is dominant, Eq. (\ref{three_body_cond_3})  becomes redundant and the three conditions reduce to two,
  \begin{equation}
  \cot \delta_{S_{ij}} (q_{E_{ij}} ) =   \mathcal{M}^{(\mathbf{ 0})}_{S_{ij} M_{S_{ij}}, S'_{ij} M'_{S_{ij}}}  (q_{E_{ij}} ), \nonumber
  \end{equation}
  and Eq. (\ref{three_body_cond_2}). Because of a finite-volume with cubic boundaries, the continuous rotation symmetry is reduced to the little group of allowed cubic rotations that leave the centre-of-mass momentum invariant,  so that,   Eq. (\ref{three_body_cond_1}),  (\ref{three_body_cond_2}) and  (\ref{three_body_cond_3}) have to be subduced according to  irreducible representations  of the appropriate little groups    \cite{Christopher_subduction:2012}.

From  Eq. (\ref{three_body_cond_1}),  (\ref{three_body_cond_2}) and  (\ref{three_body_cond_3}), we find that even in relatively simple cases,  {\it e.g.}        a single isobar dominating a single relevant partial wave, extracting the phase-shifts and determining the invariant mass $E_{ij}$ from the measured total three-particle energy $E$ is a rather difficult task. For additional information, we can first perform computations of two-particle correlators with the quantum numbers of the isobar channel to obtain information on $E_{ij}$. However, in the three-body calculation, there may be multiple $E_{ij}$ for each individual $E$ allowed by kinematics ($2 m<E_{ij}<E-m$), thus, finding the correspondence between $E_{ij}$ and $E$ may requires some assumptions and model input.   

In recent works \cite{Jo:2010,Edwards:2011}, determining the spin of the excited states by considering the overlap of carefully constructed operators with the particular state, $\langle n| \mathcal{O}| 0 \rangle$, has been proven to be successful in certain cases. We may use a similar idea to identify the $E_{ij}$ and $E$ relation in a three-particle system, by considering the overlap of operators with the state having the particular quantum numbers $(S_{ij}, L_{k}, J)$ and invariant mass of the isobar pair $E_{ij}$.  For instance, working in center-of-mass frame, the three-particle operator may be constructed in such way that  an isobar pair $(ij)$ has definite relative momentum, $\tfrac{1}{2}| \mathbf{ p}_{i}-\mathbf{ p}_{j}|= \frac{2\pi}{L} |\mathbf{ n}_{q}|, \mathbf{ n}_{q} \in \mathbb{Z}^{3}$, and definite total momentum, $|  \mathbf{ p}_{i}+\mathbf{ p}_{j} |= \frac{2\pi}{L} |\mathbf{ n}_{k}|, \mathbf{ n}_{k} \in \mathbb{Z}^{3}$. This operator will strongly overlap with the state having  invariant mass of the isobar pair $(ij)$, $E_{ij} \simeq 2 m^{2}+ \frac{1}{m} (\frac{2\pi}{L} \mathbf{ n}_{q})^{2}$, and the total energy of the three-particle system $E\simeq  m + E_{ij}+   \frac{3}{4m}(\frac{2\pi}{L} \mathbf{ n}_{k})^{2}  $ with small shifts caused by the interaction. Additionally, in the case of isobar dominance, fermion bilinear operators subduced from spin $S_{ij}$ are likely to have good overlap.

\section{Summary} \label{summary}

For the case of three non-relativisitic particles undergoing scattering, we considered in finite-volume the isobar model approximation, where quasi-two-body scattering is assumed to be dominant. Under the isobar model approximation, the  three-particle  partial wave scattering amplitudes can be  factorized as   the product of two individual scattering amplitudes, one describes the scattering of two particles inside isobar pair, and another describes the scattering between isobar pair and the  spectator particle. Three determinantal conditions, Eq. (\ref{three_body_cond_1}),  (\ref{three_body_cond_2}) and  (\ref{three_body_cond_3}),   were obtained for three-particle scattering in a finite volume.  One condition, Eq. (\ref{three_body_cond_1}),  relates  the scattering phase shifts of  two particles inside an isobar pair  to the   invariant mass of the isobar pair. The other two conditions, Eqs.  (\ref{three_body_cond_2}) and  (\ref{three_body_cond_3}),  relate the scattering phase-shifts  between an isobar pair and the spectator  to the  total energy of the three-particle system.  A proposal for extracting the phase-shifts of a three-particle system from lattice QCD simulations  is presented that makes use of carefully constructed two-particle operators within the overall three-particle operator construction.

\section{ACKNOWLEDGMENTS}
We thank  J.~J.~Dudek, R.~G.~Edwards and A.~P.~Szczepaniak  for useful discussions, and our colleagues within the Hadron Spectrum Collaboration for their  assistance.  PG acknowledges support from U.S. Department of Energy contract DE-AC05-06OR23177, under which Jefferson Science Associates, LLC, manages and operates Jefferson Laboratory.

\appendix


\section{Production amplitudes and Faddeev equations}
\label{production}   
Our task here is to consider the transition from a resonance, treated as though it were an asymptotic state, in the far past, to a system of three particles in the far future. In this case the $S$-matrix  \cite{Gellmann:1953} for the transition reads
\begin{eqnarray}
   \langle \mathbf{ p}_{1}  \mathbf{ p}_{2}   \mathbf{ p}_{3}   | U(\infty,0 )  U(0, -\infty) | R\rangle = \langle\Psi^{\mathrm{out}}(0)|\Psi^{\mathrm{in}} (0) \rangle ,  \nonumber \\
\end{eqnarray}  
where $U(t, t_{0}) =\mathcal{T}  \left \{ \exp \left [- i  \int_{t_{0}}^{t} dt'  e^{i H^{\textsc{0}}t'  }  V^{\textsc{FSI}}  e^{-i  H^{\textsc{0}} t'  }    \right ]  \right \}$ and  the incoming and outgoing wave states at $t=0$ are defined by
\begin{eqnarray}
&& |  \Psi^{\mathrm{in}} (0) \rangle  =  U(0, -\infty) | R\rangle ,   \nonumber \\
&& |\Psi^{\mathrm{out}}(0) \rangle=  U(0, \infty )  | \mathbf{ p}_{1}  \mathbf{ p}_{2}   \mathbf{ p}_{3}   \rangle   . \nonumber
\end{eqnarray}

Standard manipulations lead to the production amplitude 
\begin{eqnarray} 
   f (  \mathbf{ p}_{1}  \mathbf{ p}_{2}   \mathbf{ p}_{3}   \leftarrow R)  =  \langle  \mathbf{ p}_{1}  \mathbf{ p}_{2}   \mathbf{ p}_{3}  | V^{\textsc{FSI}}     |\Psi^{\mathrm{in}} (0) \rangle. 
\end{eqnarray} 
 The incoming wave state satisfies equation
\begin{eqnarray} 
     |  \Psi^{\mathrm{in}} (0) \rangle  =  \left [ 1+  \frac{1}{E- H^{\textsc{FSI}}  + i 0}V^{\textsc{FSI}}     \right ]  | R\rangle .   
\end{eqnarray} 
So that, the production amplitude is finally given by
\begin{eqnarray} 
  f (  \mathbf{ p}_{1}  \mathbf{ p}_{2}   \mathbf{ p}_{3}   \leftarrow R)  =  \langle  \mathbf{ p}_{1}  \mathbf{ p}_{2}   \mathbf{ p}_{3}  |\,  T^{\textsc{FSI}} \,   |R \rangle, 
\end{eqnarray} 
where the $T$-matrix operator is defined by 
\begin{eqnarray}
T^{\textsc{FSI}}  = V^{\textsc{FSI}} +  V^{\textsc{FSI}}  \frac{1}{E- H^{\textsc{FSI}} + i \epsilon}  V^{\textsc{FSI}}  . \nonumber \\
\end{eqnarray}
Within the isobar model we assume dominance of pair-wise interactions, and thus the $T$-matrix for the three-particle system can be decomposed as  $T^{\textsc{FSI}}  = \sum_{k} T^{(ij)k}$, where $T^{(ij)k}$ satisfies Faddeev equations \cite{Faddeev:1960},
\begin{eqnarray}
T^{(ij) k} = t^{(ij)} + t^{(ij)} \frac{1}{E-H^{\textsc{0}} + i \epsilon} \left [ T^{(jk) i} + T^{(ki ) j} \right ], \nonumber \\
\end{eqnarray}
where $t^{(ij)}$ is the two-body $T$-matrix for pair $(ij)$. 

\begin{widetext}
It follows that the three-particle production amplitude can be decomposed as  \mbox{$f (  \mathbf{ p}_{1}  \mathbf{ p}_{2}   \mathbf{ p}_{3}   \leftarrow R) = \sum_{k} f^{(ij)k} (  \mathbf{ q}_{ij}     \, \mathbf{ k}_{k}   \leftarrow R)$, where  $ f^{(ij)k} (  \mathbf{ q}_{ij}    \,  \mathbf{ k}_{k}   \leftarrow R)$} satisfies an integral  equation  \cite{Ahmadzadeh:1965}
\begin{eqnarray}\label{faddeeveq} 
 f^{(ij)k} (  \mathbf{ q}_{ij}   \,   \mathbf{ k}_{k}   \leftarrow R) &=&  \langle  \mathbf{ q}_{ij}    \,  \mathbf{ k}_{k} | t^{(ij)}    |R \rangle \nonumber \\
 &\quad\quad+&\int  \frac{d^{3} \mathbf{ q}'_{jk}    }{(2\pi)^{3}}    \frac{d^{3} \mathbf{ k}'_{i}    }{(2\pi)^{3}}   \frac{   \langle  \mathbf{ q}_{ij}  \,   \mathbf{ k}_{k} | t^{(ij)}   |\mathbf{ q}'_{jk}      \mathbf{ k}'_{i}    \rangle  \;  f^{(jk)i} (  \mathbf{ q}'_{jk}    \,  \mathbf{ k}'_{i}   \leftarrow R)   }{E-H^{\textsc{0}} ( \mathbf{ q}'_{jk}  \,     \mathbf{ k}'_{i}  ) + i \epsilon}   \nonumber \\
 &\quad\quad+ &\int  \frac{d^{3} \mathbf{ q}'_{ki}    }{(2\pi)^{3}}    \frac{d^{3} \mathbf{ k}'_{j}    }{(2\pi)^{3}}   \frac{ \langle  \mathbf{ q}_{ij}   \,    \mathbf{ k}_{k} | t^{(ij)}   |\mathbf{ q}'_{ki}    \,  \mathbf{ k}'_{j}  \rangle   \;  f^{(ki)j} (  \mathbf{ q}'_{ki}    \,   \mathbf{ k}'_{j}   \leftarrow R)  }{E-H^{\textsc{0}} ( \mathbf{ q}'_{ki}  \,     \mathbf{ k}'_{j}  ) + i \epsilon}.  \nonumber \\
 \end{eqnarray} 
 \end{widetext}
The first term on the right hand side in Eq. (\ref{faddeeveq}) is the Born-term for production of the isobar pair $(ij)$ plus the $k^\mathrm{th}$ particle from the resonance state $| R\rangle$. The remaining terms generate the rescattering effect from different isobar pairs -- the isobar model approximation is to keep only the Born term in Eq. (\ref{faddeeveq}), so that the production amplitude is approximated by
\begin{eqnarray} 
   f (  \mathbf{ p}_{1}  \mathbf{ p}_{2}   \mathbf{ p}_{3}   \leftarrow R)   \approx  \sum_{k=1,2,3}^{ i \neq j \neq k} \langle  \mathbf{ q}_{ij}      \mathbf{ k}_{k} | t^{(ij)}    |R \rangle. 
\end{eqnarray}


\begin{thebibliography}{99}


\bibitem{Michael:1985ne} 
  C.~Michael,
  Nucl.\ Phys.\ B {\bf 259}, 58 (1985).

\bibitem{Luscher:1990ck} 
  M.~Luscher and U.~Wolff,
  Nucl.\ Phys.\ B {\bf 339}, 222 (1990).

\bibitem{Blossier:2009kd} 
  B.~Blossier, M.~Della Morte, G.~von Hippel, T.~Mendes and R.~Sommer,
  JHEP {\bf 0904}, 094 (2009)
  
  

      
      
  \bibitem{Jo:2010}
	J. J. Dudek  {\it et al.}  (Hadron Spectrum Collaboration),
  Phys.\ Rev.\ D {\bf 82}, 034508 (2010).

  \bibitem{Edwards:2011}
	R. G. Edwards, J. J. Dudek, D. G. Richards, and S. J. Wallace,
  Phys.\ Rev.\ D {\bf 84}, 074508 (2011).

  \bibitem{Jo_scatt:2012}
	J. J. Dudek  {\it et al.}  (Hadron Spectrum Collaboration),
	Phys.\ Rev.\ D {\bf 86}, 034031 (2012).






       
       

  \bibitem{Feng:2011}
	X. Feng, K. Jansen, and D. B. Renner,
  Phys.\ Rev.\ D {\bf 83}, 094505 (2011).

  \bibitem{Jo_scatt:2011}
	J. J. Dudek  {\it et al.}  (Hadron Spectrum Collaboration),
  Phys.\ Rev.\ D {\bf 83}, 071504 (2011).

  \bibitem{Beane_scatt:2012}
	S. R. Beane  {\it et al.}  (NPLQCD Collaboration),
	Phys.\ Rev.\ D {\bf 85}, 034505 (2012).


  \bibitem{Lang_scatt:2011}
	C. B. Lang, D. Mohler, S. Prelovsek and M. Vidmar,
	Phys.\ Rev.\ D {\bf 84}, 054503 (2011).

\bibitem{Aoki:2011yj} 
  S.~Aoki {\it et al.}  [CS Collaboration],
  Phys.\ Rev.\ D {\bf 84}, 094505 (2011)
  
  
  
  

  
  \bibitem{Dudek:2012xn} 
  J.~J.~Dudek, R.~G.~Edwards and C.~E.~Thomas,
  arXiv:1212.0830 [hep-ph].




\bibitem{Maiani:1990ca} 
  L.~Maiani and M.~Testa,
  Phys.\ Lett.\ B {\bf 245}, 585 (1990).



  

  \bibitem{Lusher:1991}
M. L\"uscher,
  Nucl.\ Phys.\ B {\bf 354}, 531 (1991).



  \bibitem{Gottlieb:1995}
K. Rummukainen, S. Gottlieb,
  Nucl.\ Phys.\ B {\bf 450}, 397 (1995).

  \bibitem{Lin:2001}
C.-J.D. Lin, G. Martinelli, C. T. Sachrajda and M. Testa,
  Nucl.\ Phys.\ B {\bf 619}, 467 (2001).
            
  \bibitem{Christ:2005}
	N. H. Christ, C. Kim  and T.Yamazaki,
  Phys.\ Rev.\ D {\bf 72}, 114506 (2005).
            
  \bibitem{Bernard:2007}
V. Bernard, Ulf-G. Mei{\ss}ner and A. Rusetsky,
  Nucl.\ Phys.\ B {\bf 788}, 1 (2008).
  
  \bibitem{Bernard:2008}
V. Bernard, M. Lage, Ulf-G. Mei{\ss}ner and A.Rusetsky,
  JHEP {\bf 0808}, 024 (2008).
  
   
  \bibitem{Liu:2005}
S. He, X. Feng, C. Liu,
  JHEP {\bf 0507}, 011 (2005).


  \bibitem{Doring:2011}
M. D\"oring, Ulf-G. Mei{\ss}ner,E. Oset and A. Rusetsky,
Eur.\ Phys.\ J.\ A {\bf 47}, 139 (2011)




\bibitem{Aoki:2011} 
  S.~Aoki {\it et al.}  [HAL QCD Collaboration],
  Proc.\ Japan Acad.\ B {\bf 87}, 509 (2011)
  
\bibitem{Briceno:2012yi} 
  R.~A.~Briceno and Z.~Davoudi,
  arXiv:1204.1110 [hep-lat].


\bibitem{Hansen:2012tf} 
  M.~T.~Hansen and S.~R.~Sharpe,
  Phys.\ Rev.\ D {\bf 86}, 016007 (2012)








\bibitem{Guo:2013cp} 
P.~Guo, J.~Dudek, R.~Edwards, A.~P.~Szczepaniak,
  [arXiv:1211.0929  [hep-lat]].









  \bibitem{Goradia:1977}
	Y. Goradia  and T. A. Lasinski,
  Phys.\ Rev.\ D {\bf 15}, 220 (1977).
  
  \bibitem{Ascoli:1975}
	G. Ascoli  and H. W. Wyld,
  Phys.\ Rev.\ D {\bf 12}, 43 (1975).



  
    
    
  
  
  



  
  
  
  




  \bibitem{Faddeev:1965}
	L. D. Faddeev,
  {\it Mathematical Aspects of the Three-Body Problem in the Quantum Scattering Theory} (Israel Program for Scientific Translation, Jerusalem, Israel, 1965).




  \bibitem{Faddeev:1960}
	L. D. Faddeev,
  Zh.\ Eksp. \ Teor.\ Fiz. \  {\bf 39}, 1459 (1960) [Sov. Phys.-JETP {\bf 12}, 1014(1961)]. 




  \bibitem{Rusetsky:2012}
K. Polejaeva and A. Rusetsky,
  Eur.\ Phys.\ J.\ A{\bf 48}, 67 (2012).




  \bibitem{Briceno:2012}
R. A. Briceno and Z. Davoudi,
 [arXiv:1212.3398  [hep-lat]].





  \bibitem{Hwa:1963}
	 R. C. Hwa,
  Phys.\ Rev. {\bf 130}, 2580 (1963).




  \bibitem{Aaron:1973}
	R. Aaron and R. D. Amado,
  Phys.\ Rev. Lett.\  {\bf 31}, 1157 (1973).



 




  \bibitem{Christopher_subduction:2012}
	C. E. Thomas, R. G. Edwards  and J. J. Dudek,
  Phys.\ Rev.\ D {\bf 85}, 014507 (2012).
  







 

















  \bibitem{Gellmann:1953}
  M. Gell-Mann and M. L. Goldberger,
  Phys.\ Rev.  {\bf 91}, 398 (1953).



  \bibitem{Ahmadzadeh:1965}
  A. Ahmadzadeh and J. A. Tjon,
  Phys.\ Rev.  {\bf 139}, B1085 (1965).




    
\end{thebibliography}
\end{document}